# Nanoparticle Unification of Dark Matter and Dark Energy? Cancellation of the Vacuum Catastrophe

**Keith Johnson [1]***

[1] Massachusetts Institute of Technology, Cambridge, MA 02139
* Correspondence: kjohnson@mit.edu

**Abstract:** Laboratory creation of stable "magic-number" protonated water nanoclusters from amorphous water-ice under simulated astrophysical conditions suggests the possible ejection of such nanoparticles from ubiquitous amorphous water-ice-covered cosmic dust to interstellar space. The quantum properties of these quintessential nanoparticles introduce a tantalizing prospect that bridges the origins of dark matter and dark energy. The quantum-entangled diffuse Rydberg electronic states inherent to cosmic water nanoclusters make them plausible candidates for baryonic dark matter. Moreover, they exhibit the capacity to absorb, through the microscopic dynamical Casimir effect, virtual photons originating from zero-point-energy vacuum fluctuations occurring above the water-nanocluster vibrational frequency cutoff. This selective interaction leaves only vacuum fluctuations below these frequencies with gravitational significance, cancelling the vacuum-energy catastrophe and yielding a shared genesis for dark matter and dark energy. These findings coalesce into a cosmological framework depicting a cyclic universe, with cosmic water nanoclusters constituting a quintessence scalar field, instead of adhering to the multiverse concept based on cosmic inflation theory. Recent observations of CMB birefringence support this quintessence.

## 1. Introduction

In the standard framework of Big Bang cosmology, the matter and energy composition of our universe is governed by two enigmatic entities: *dark matter*, a nonbaryonic substance with an unknown nature, and *dark energy*, a strange negative-pressure field. Despite over three decades of searching, proposed exotic elementary particles such as weakly interacting massive particles (WIMPs) and AXIONS have yet to materialize in experimental observations. Similarly, predictions of WIMPs stemming from supersymmetry theory have not been realized in the CERN Large Hadron Collider. Dark energy, responsible for the accelerating expansion of the universe, is typically regarded as a distinct challenge from dark matter and is associated with fluctuations of the zero-point energy of the cosmic vacuum. Moreover, quantum field theory predicts a vacuum energy density that exceeds the observed value by as much as a staggering factor of $10^{120}$, known as the *cosmological constant problem* [1]. In this paper, I propose that water nanoclusters, ejected from amorphous water-ice enveloping ubiquitous cosmic dust, as simulated in the laboratory [2], when excited to their diffuse Rydberg electronic states, may serve as plausible candidates for *baryonic* dark matter [3]. The cutoff terahertz (THz) vibrational frequencies of these water nanoparticles align closely with the $\nu_c \cong 1.7$ THz cutoff frequency of vacuum fluctuations, proposed by Beck and Mackey [4] to account for the small values of vacuum energy and the cosmological constant. I suggest that cosmic water nanoclusters can also harness high-frequency vacuum zero-point-energy virtual photons through the microscopic *dynamical Casimir effect* [5], leaving only the low-frequency ones to exert gravitational influence, thereby cancelling the infamous *vacuum-energy catastrophe* or cosmological constant problem [1]. Finally, I propose that cosmic water nanoclusters, ejected from cosmic dust and other water-ice-coated astrophysical sources, distributed as a low-density "dark gas or fluid" throughout the universe [6,7], might represent a plausible shared source of both dark matter and dark energy, alongside or instead of yet-to-be-discovered exotic elementary particles such as WIMPS and AXIONS.

## 2. Cosmic Water Nanoclusters: Electronic Structure and Terahertz (THz) Vibrations

The continued non-detection of WIMP and AXION elementary particles, which have been prime candidates for dark matter, raises a fundamental question: Can cosmic water nanoparticles hold clues to understanding these cosmic enigmas? It is essential to recognize the predominant role of hydrogen and oxygen - the constituents of water - in our universe. Both are the most abundant and chemically active. Intriguingly, water vapor plays a pivotal role during the early phases of star creation. In these regions, water acts as a significant oxygen reservoir, ensuring the effective cooling of the surrounding gas, which is a crucial process in star evolution [8]. In fact, regions such as the Orion nebula are known to generate immense amounts of water daily, much more than the volume of Earth's oceans [9]. Moreover, the discovery of massive reservoirs of water, associated with high-redshift quasars, twelve or more billion light-years from Earth, underscores the widespread presence of water in the universe [10]. Such quasars, housing astounding masses of water vapor, are indicative of water's abundance even in remote cosmic locales. Furthermore, recent studies suggest that during the universe's infancy - the initial billions of years post Big Bang - water could have been quite widespread [11]. Closer to home, stable water nanoclusters arise in the atmosphere because of hydrogen bonding between water molecules [12]. A plausible mechanism for the formation of these nanoclusters in the cosmos is cosmic-ray-induced direct ejection from amorphous ice covering cosmic dust grains, as observed under laboratory conditions [2]. These grains, which are prevalent in interstellar clouds, owe their existence to supernovae explosions [13]. The cosmic-ray-induced ionization of $H_2$ molecules adsorbed on amorphous water-ice, namely, $H_2^+ + nH_2O + \text{grain} \rightarrow H_3O^+(H_2O)_{n-1}\uparrow + \text{grain}$, is another proposed scenario for the ejection of protonated water nanoclusters to interstellar space [14]. These ionized water nanoclusters, owing to their oscillating electric dipole moments, are linked to the observed THz emissions from water vapor under intense UV radiation [15] and thus should be stable under similar cosmic radiation. Interestingly, $H_3O^+(H_2O)_{20}$ or its equivalent protonated form, $(H_2O)_{21}H^+$ is notably stable under vacuum conditions. This "magic-number" structure can be visualized as a hydronium ion, $H_3O^+$ encased by a pentagonal dodecahedral assembly of twenty water molecules (Figure 1d). The stability of this unique structure may have cosmic implications, as suggested by the recent observation of interstellar hydronium [16]. Spectroscopic identification of larger protonated cosmic water nanoclusters is a challenge, but crucial for confirming their presence and understanding their possible role in cosmic phenomena.

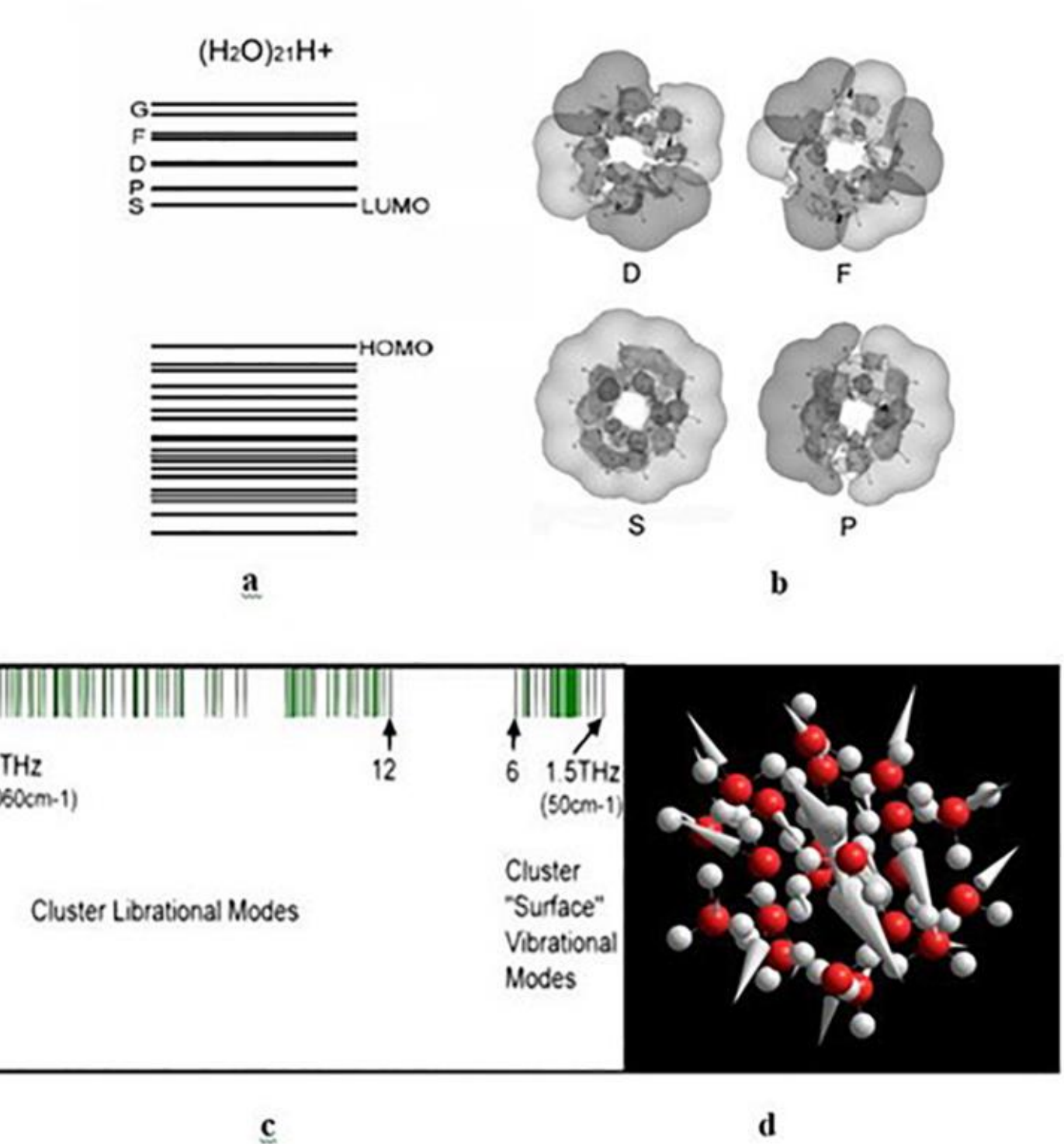

**Figure 1.** Molecular-orbital energy levels, wavefunctions, and vibrational modes of the protonated $(H_2O)_{21}H^+$ cluster, calculated using the SCF-X$\alpha$-Scattered-Wave density-functional method (Reference 17). **a.** Cluster molecular-orbital energy levels. The HOMO-LUMO energy gap is approximately 3 eV. **b.** Wavefunctions of the lowest unoccupied cluster molecular orbitals. **c.** THz vibrational spectrum. **d.** Lowest-frequency THz vibrational mode and relative vibrational amplitudes. The vectors show the directions and relative amplitudes for the oscillation of the hydronium $(H_3O)^+$ oxygen atom coupled to the O–O–O "bending" motions of the cluster "surface" oxygen atoms.

Figure 1 illustrates the ground-state molecular-orbital energies, wavefunctions, and vibrational modes of the protonated pentagonal dodecahedral water cluster, $(H_2O)_{21}H^+$. These results were obtained using the SCF-X$\alpha$-Scattered-Wave density-functional method, a collaborative development by the author [17]. Molecular dynamics simulations confirm the stability of the cluster at temperatures exceeding 100°C, with minimal changes. Similar calculations were performed for the neutral pentagonal dodecahedral water cluster, $(H_2O)_{20}$, and its arrays, revealing THz vibrational modes, as depicted in Figures. 2 and 3.

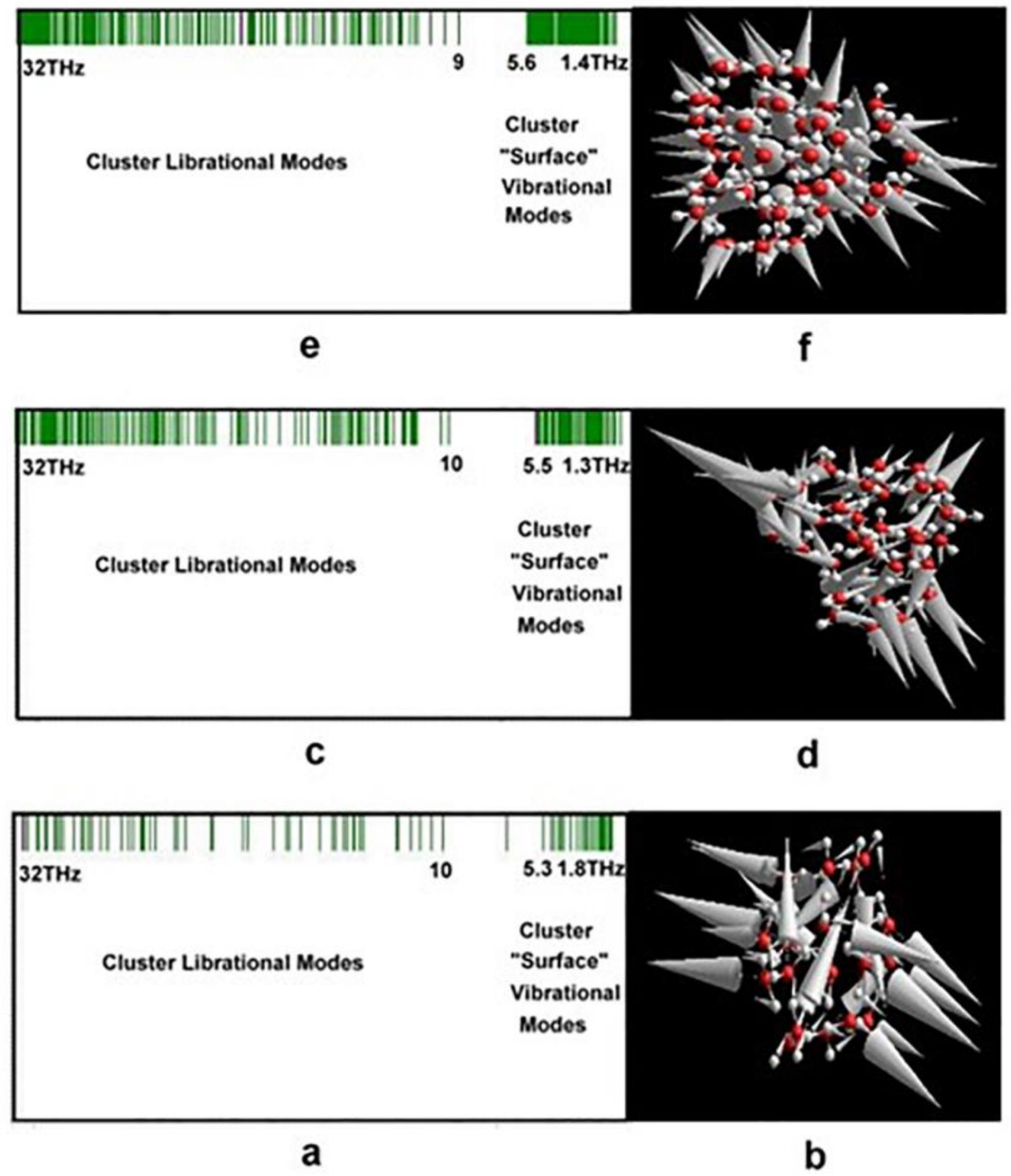

**Figure 2. a.** THz vibrational spectrum of a pentagonal dodecahedral $(H_2O)_{20}$ cluster calculated using the SCF-X$\alpha$-Scattered-Wave density-functional method (Reference 17). **b.** Lowest-frequency THz vibrational mode and relative vibrational amplitudes. **c.** Calculated THz vibrational spectrum of a stable array of three dodecahedral water clusters. **d**. Lowest-frequency THz vibrational mode. **e.** THz vibrational spectrum of a stable array of five dodecahedral water clusters. **f.** Lowest-frequency THz vibrational mode.

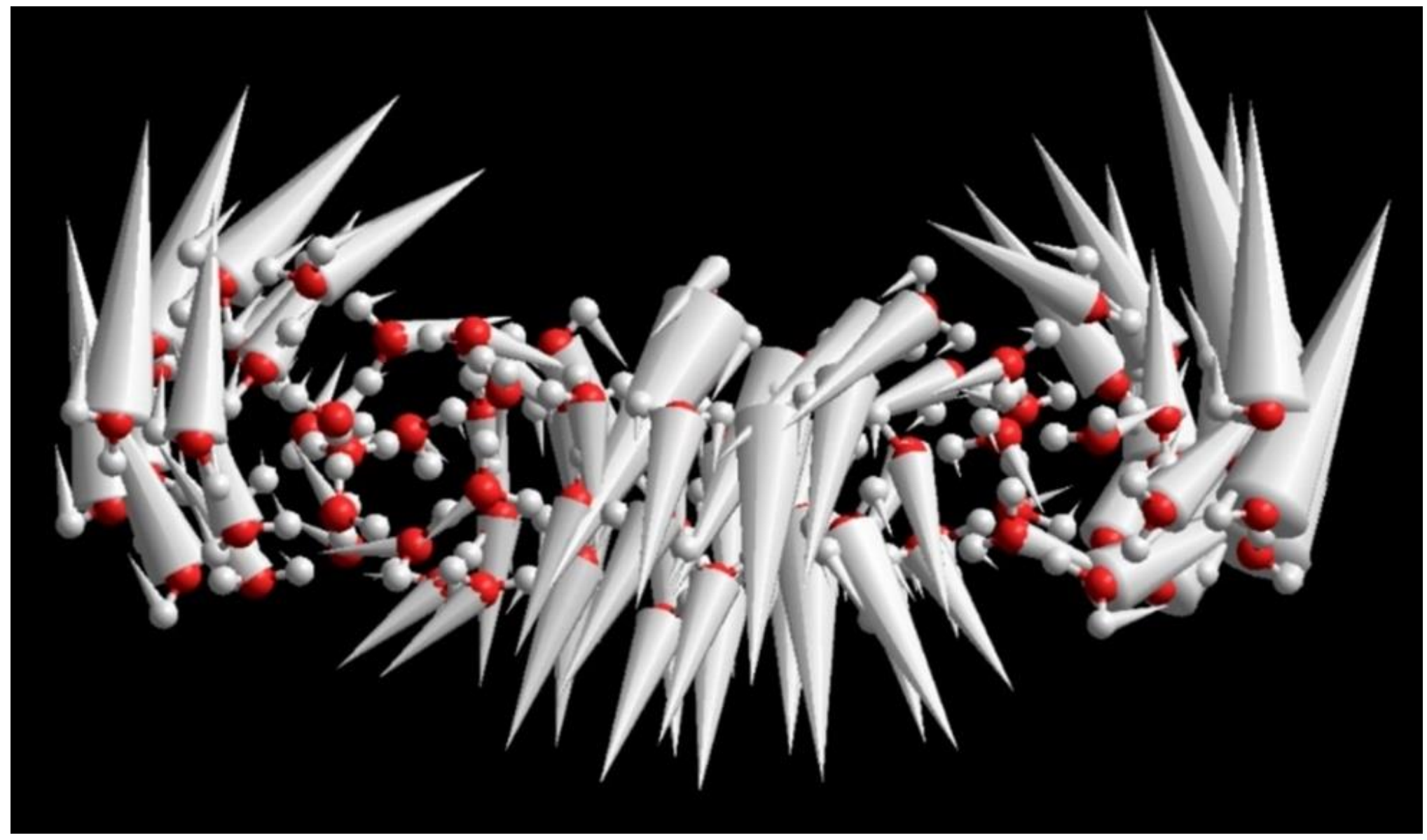

**b**

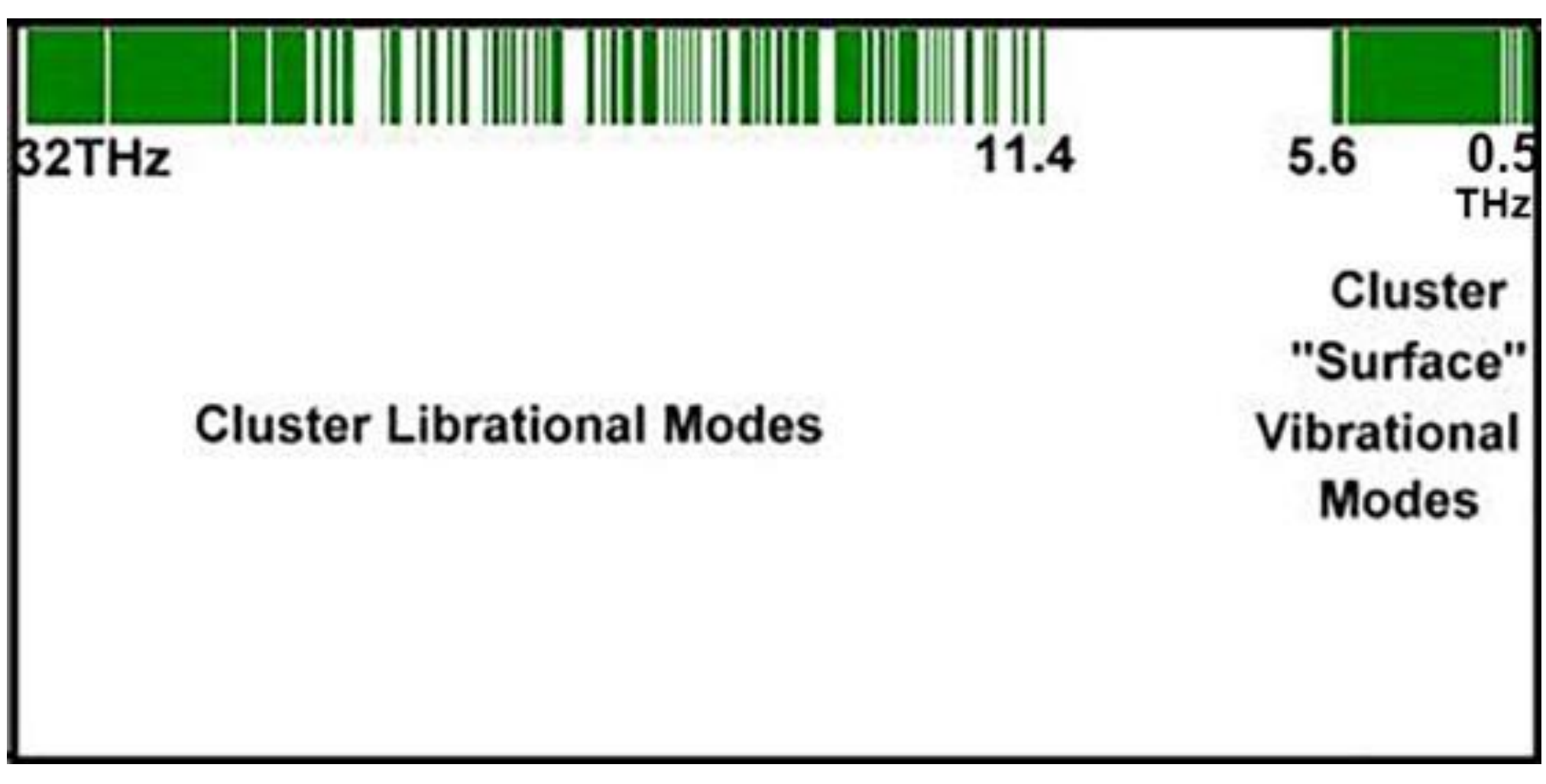

**a**

**Figure 3. a.** THz vibrational spectrum of a stable linear array of five pentagonal dodecahedral water nanoclusters calculated using the SCF-X$\alpha$-Scattered-Wave density-functional method (Reference 17). **b.** Lowest-frequency THz vibrational mode and relative vibrational amplitudes. Note the lower cutoff THz frequency compared to those in Figures 2e and f.

Figures. 2 and 3 exhibit qualitative resemblance to Figure 1c,d but show a gradual decrease in the vibrational frequency cutoff as the cluster size increases. This trend aligns with experimental observations of THz radiation emission from water vapor nanoclusters [15], as depicted in Figure. 4. The shift in THz emission peaks towards lower frequencies and intensities, corresponding to larger clusters, occurs with decreasing water vapor ejection pressure in the vacuum chamber where the radiation was measured. This observation suggests a decrease in THz emission cutoff frequencies and intensities as water nanoclusters of increasing sizes are ejected from ice-coated cosmic dust.

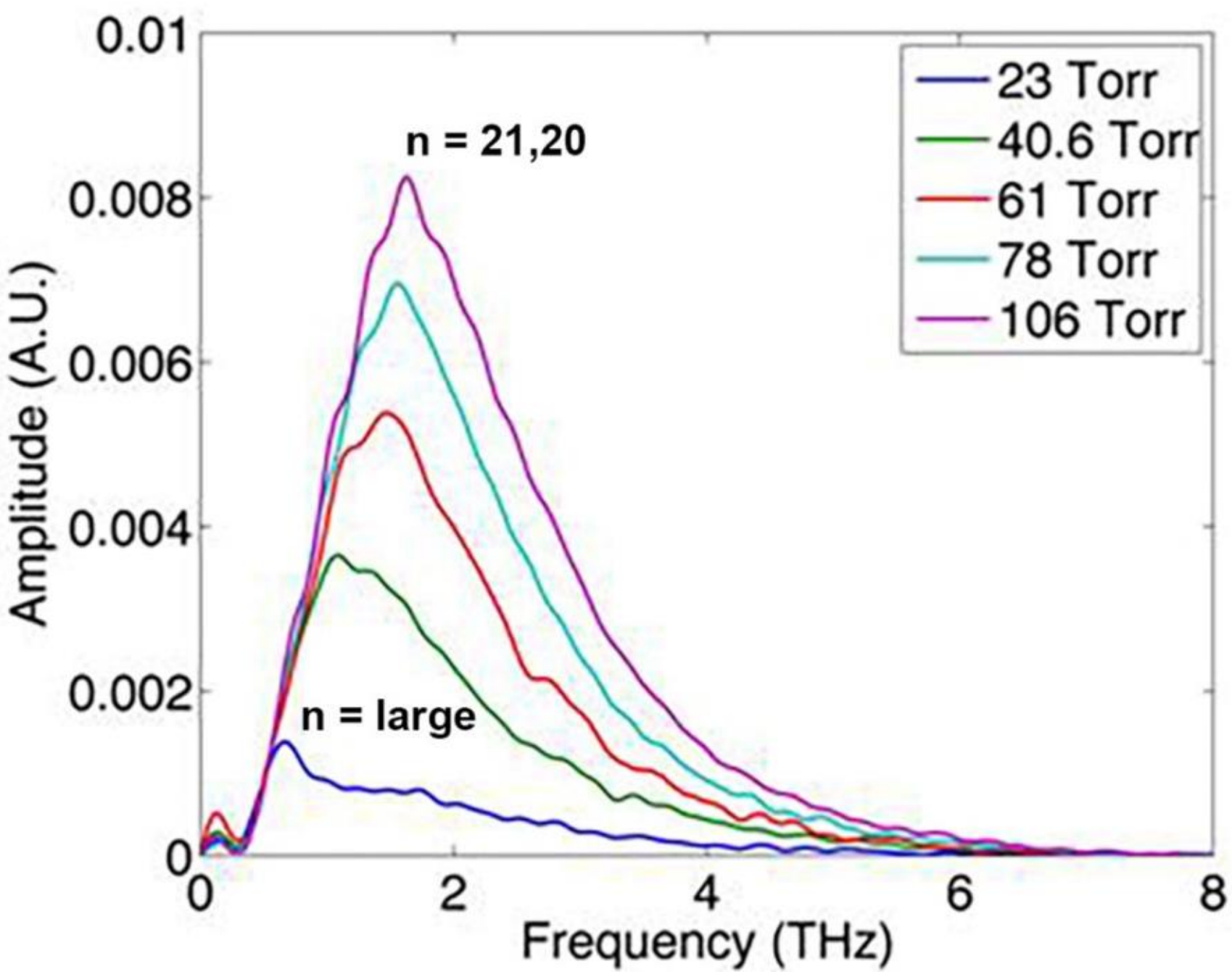

**Figure 4.** Frequency dependence for a range of pressures of the THz wave generation amplitudes (in arbitrary units) from water nanoclusters, $(H_2O)_nH^+$ and $(H_2O)_n$ produced from water vapor, described in Ref. 15. Maximum wave generation is associated with the “magic numbers,” n = 20 and 21.

Common characteristics among all these water clusters include: (1) lowest unoccupied (LUMO) energy levels, akin to those in Figure 1a, corresponding to diffuse Rydberg cluster "surface" molecular-orbital wavefunctions depicted in Figure 1b as "S," "P," "D," and "F"-like orbitals and (2) bands of vibrational modes ranging from 0.5 to 6 THz (Figures 1-3) due to O-O-O "squashing" or "bending" and "twisting" motions between adjacent hydrogen bonds. The vectors in Figures 1-3 represent the directions and relative amplitudes of the lowest THz-frequency modes associated with the O-O-O "bending" (or "squashing") motions of the water-cluster "surface" oxygen atoms. Observations of surface O-O-O bending vibrations in this energy range have been performed under laboratory conditions [18]. Furthermore, ultraviolet excitation of an electron from the highest occupied molecular orbital (HOMO) to the lowest unoccupied molecular orbital (LUMO), as shown in Figure 1a, can place the electron into the Rydberg "S"-like cluster molecular orbital illustrated in Figure 1b. Occupancy of this orbital results in a bound state, even when an additional electron is introduced, leading to the formation of a "hydrated electron" [19]. In contrast, a water monomer or dimer exhibits virtually no electron affinity. Therefore, in space, particularly within dense interstellar clouds, a water nanocluster, $(H_2O)_{21}H^+$ or $H_3O^+(H_2O)_{20}$, ejected from amorphous water-ice-coated cosmic dust, is likely to capture an electron, leading to electrically neutral water nanoclusters, as shown in Figure 2.

## 3. Rydberg Dark Matter

Starlight energy can trigger electronic excitations from the $(H_2O)_{21}H^+$ HOMO (Figure. 1a) or the LUMO when a hydrated electron is captured. These excitations then shift to the more delocalized "P", "D", "F", and other higher water-cluster Rydberg orbitals shown in Figure 1b. These specific states exhibit minimal spatial overlap with lower-energy-filled states. Their longevity increases with increasing excitation energy and principal Rydberg quantum number. Thus, they become potential contenders for Rydberg Matter (RM) — a dilute phase of weakly bonded individual molecules in Rydberg excitation, which have extensive effective interactions [20]. RM is a *low-density* condensed phase of weakly interacting individual Rydberg-excited molecules over long-range intermolecular distances - approximately one micron for the $(H_2O)_{21}H^+$ Rydberg level, $n = 80$. RM is transparent at visible, infrared, and microwave frequencies and thus, ignoring the observationally difficult emissive THz region (Figure 4), qualifies as dark matter.

*Quantum entanglement* of RM occurs because the diffuse nanocluster Rydberg molecular orbitals "overlap" even at relatively large distances. Pentagonal dodecahedral water nanoclusters, containing an excited or "hydrated" electron in the Rydberg "S" LUMO, resemble oversized hydrogen atoms (Figure 1b). When two such clusters are in close proximity, they can create an interlocking "Sσ-bonding" molecular orbital as depicted in Figure 5a, which accommodates two electrons with paired spins, mirroring a large hydrogen molecule. Elevating the electrons to the Rydberg "P" LUMO results in the formation of "Pσ" and "Pπ" molecular bonding orbitals, as shown in Figures 5b and 5c. However, it's uncommon for two water nanoclusters to come near each other in the interstellar regions because of their sparse distribution. At more significant distances between these clusters, the expansive molecular orbitals of their highest, most delocalized Rydberg states can overlap enough to allow quantum entanglement across vast spatial expanses.

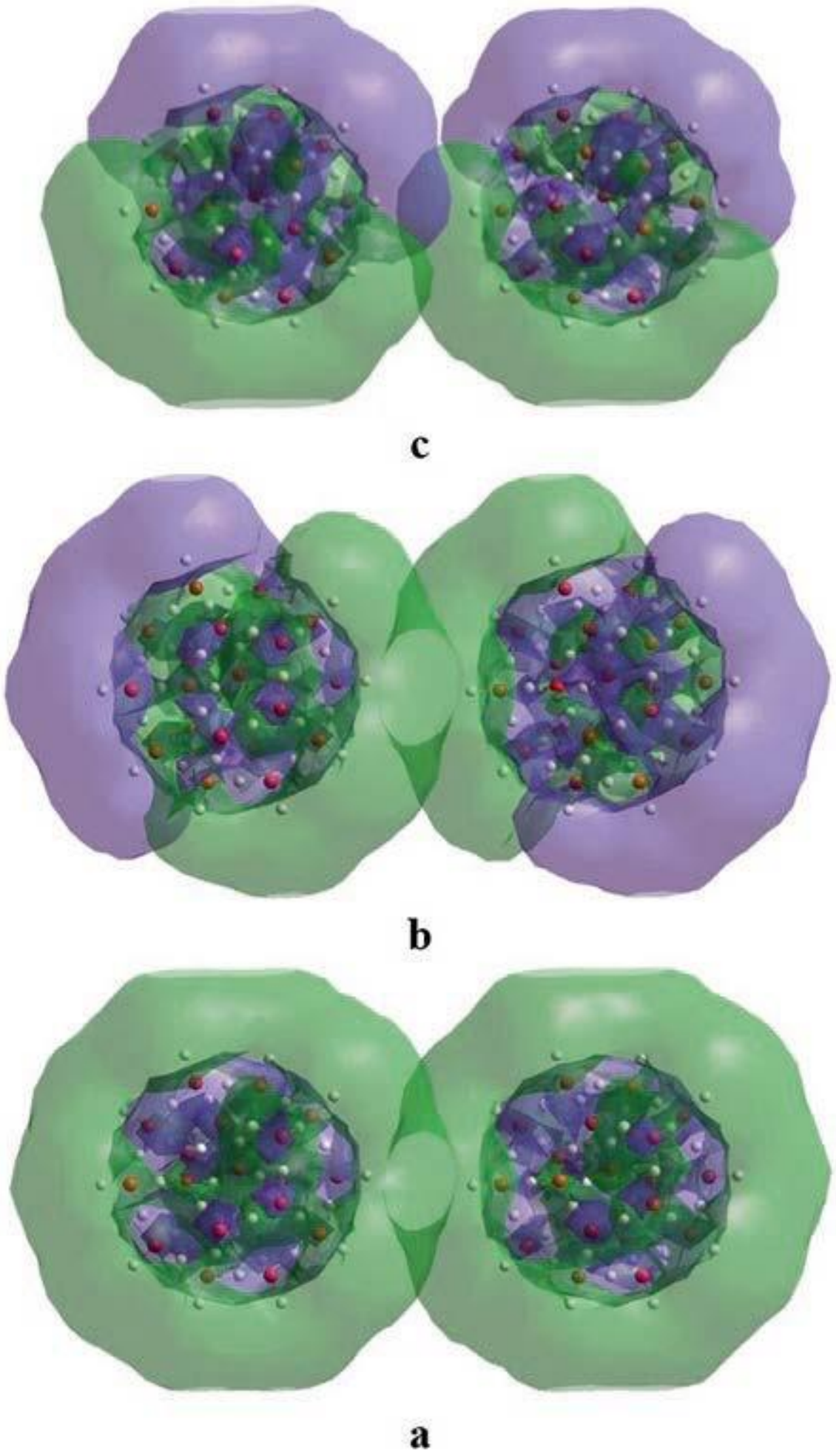

**Figure. 5. a**. “Sigma-bond” molecular-orbital overlap of water-nanocluster Rydberg “S” LUMOs shown in Figure 1b. **b**. “Sigma-bond” molecular-orbital overlap of Rydberg “P” LUMOs. **c**. “Pi-bond” molecular-orbital overlap of Rydberg “P” LUMOs.

## 4. Observational Support

### 4.1 No Cosmic Dust – No Dark Matter

The Hubble Space Telescope has revealed a galaxy, NGC 1052-DF2, approximately 72 million light years from Earth, where one can literally see through other galaxies behind it [21] (Figure 6). It is an ultra-diffuse galaxy, almost as wide as the Milky Way, but contains only one 200th of the number of stars in the Milky Way. Furthermore, the galaxy contains at most only one four hundredth of the amount of dark matter that astronomers expect from the standard dark matter theory. The ultra diffuseness of the galaxy also suggests a lack of cosmic dust characteristic of the Milky Way and most other galaxies. Thus, the proposal made in this paper that cosmic dust, coated with amorphous water-ice, produces water nanoclusters that constitute a type of baryonic dark matter is supported by the observation of galaxies such as NGC 1052-DF2, which are devoid of both cosmic dust and dark matter.

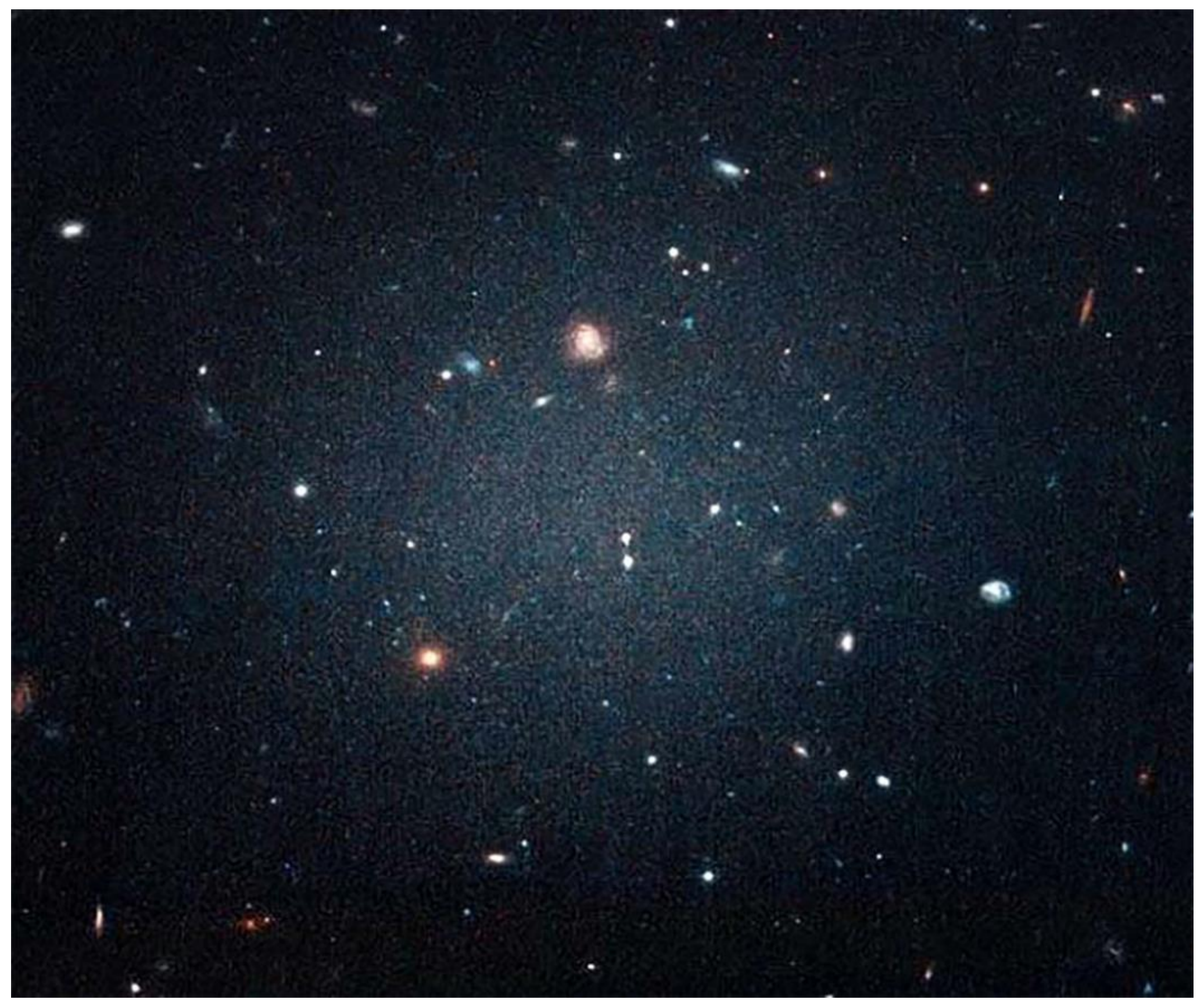

**Figure 6.** Galaxy NGC 1052-DF2. See Reference 21.

**4.2 Cosmic Birefringence**

Minami and Komatsu [22] have reported the extraction of cosmic birefringence data from the 2018 Planck Cosmic Microwave Background, which may possibly support "new physics" such as *quintessence*. A more recent publication [23] agrees with [22] and claims to be more precise than the previous paper, providing further evidence for physics beyond the standard model. The THz vibrations of cosmic water nanoclusters, ejected from amorphous water-ice-coated cosmic dust to interstellar space, produce the dipole-moment anisotropy necessary and sufficient for their birefringence property. Figure 7a illustrates the 1.8 THz vibrational mode of the cosmic water nanocluster $(H_2O)_{21}H$ which contains a hydrated electron. Dynamic Jahn-Teller symmetry breaking of the water-cluster dodecahedra produces the dipole moments along the nanocluster axes indicated in Figure 7a, computed by the SCF-X$\alpha$ cluster density-functional method [17]. Such anisotropic dipole moments lead directly to the birefringence of water nanoclusters, as they do for the observed THz-induced birefringence of ordinary liquid water [24], which can be viewed as a dynamical random network of dodecahedral water clusters. Thus, the birefringence of cosmic water nanoclusters can explain the CMB birefringence from a condensed-matter point of view, instead of by unobserved elementary particles such as AXIONS [22,23].

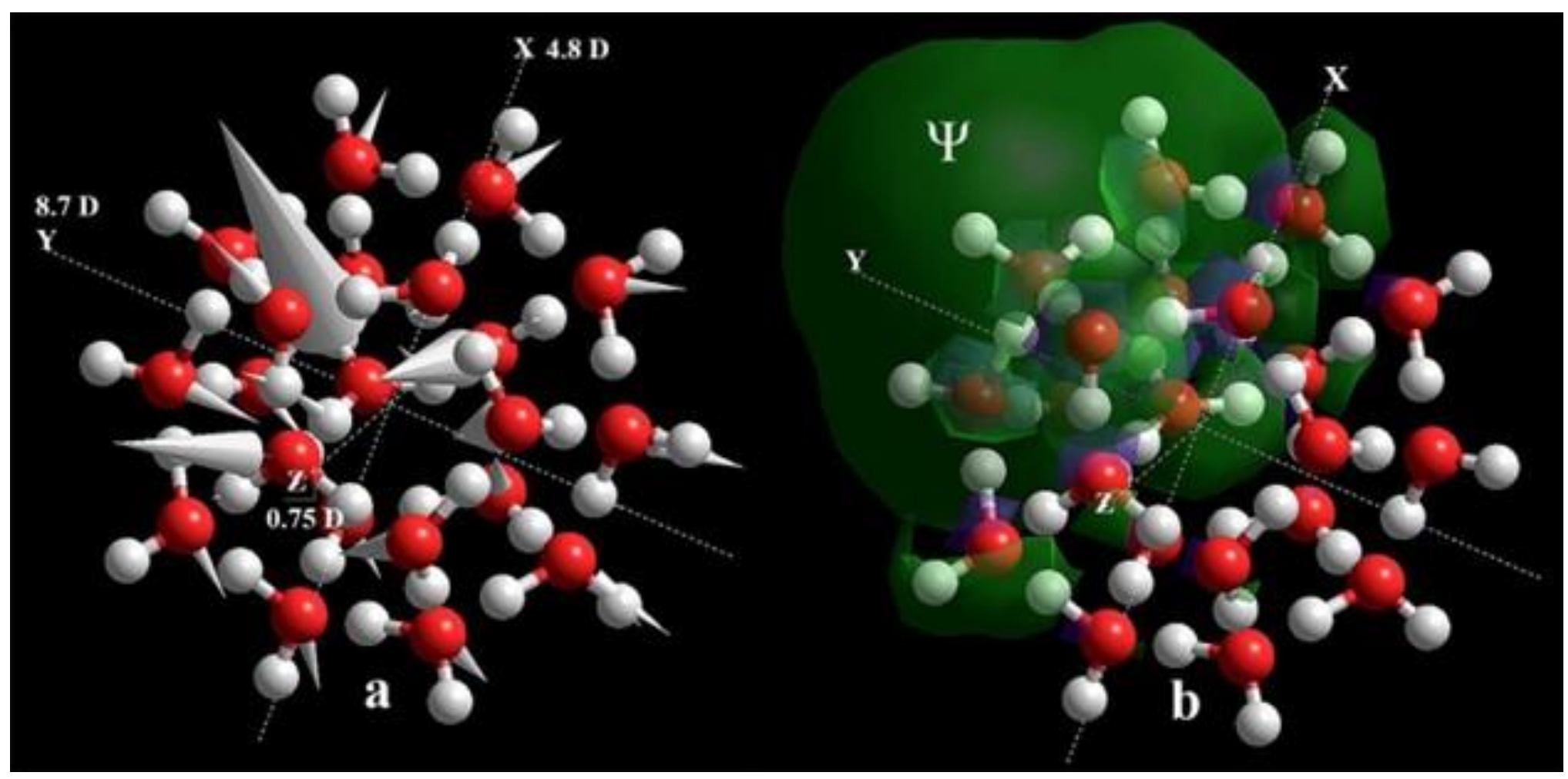

**Figure 7. a.** 1.8 THz vibrational mode of water nanocluster $(H_2O)_{21}H$ holding a hydrated electron. The component anisotropic dipole moments (in Debyes) along the nanocluster axes are shown**. b.** Wavefunction Ψ of the attached electron responsible for the anisotropic dipole moment.

### 4.3 The Bullet Cluster and Galactic Halos

Standard cosmology largely agrees that dark matter is not baryonic. One of the most compelling claims of evidence for the existence of dark matter is the gravitational lensing observations from the *Bullet Cluster*, which distinctly shows the separation between luminous matter and dark matter [25]. The protonated water nanoclusters, shown in Figure 1, are positively charged. However, it's suggested that these clusters might obtain a "hydrated" electron from space once they're ejected from cosmic dust that's ice-coated, transitioning into electrically neutral clusters like those in Figure 2. While it is generally accepted that dark matter is electrically neutral, recent discussions suggest that a small amount of charged dark matter might have played a role in cooling early universe baryons [26]. The dark matter properties of the Bullet Cluster might be linked to the electric charge of the ice-coated cosmic dust, which is believed to be the birthplace of cosmic water nanoclusters. Given that a significant portion of the Bullet Cluster's regular matter could be positively charged cosmic dust, this could amplify the ejection of protonated water nanoclusters, further highlighting the observed separation between luminous and dark matter. Even with existing uncertainties about galactic-scale electric fields, it is conceivable that these fields might promote the aggregation of cosmic water nanocluster RM around galaxy outskirts, possibly explaining the enigmatic galactic *dark matter halos*, akin to the dark matter observed in the Bullet Cluster. Despite uncertainties about electric fields on the galactic scale, it is possible that such fields over time can cause the expulsion of water-nanocluster RM into the outer dark-matter halo reaches of the galaxies, thereby explaining the recent observation [27] of a *direct interaction*, beyond gravitational, between dark matter and normal baryonic matter in galaxies. Such a direct interaction is consistent with the baryonic nature of cosmic water-nanocluster Rydberg dark matter. This creates spherical regions of relatively constant density within the dark matter halo, with dimensions that increase proportionately over time and finally reach those of the galactic stellar disc.

## 5. Dark Energy

### 5.1. The Cosmological Constant Problem

Quantum field theory proposes a vacuum energy density that exceeds the measured value by as much as 120 orders of magnitude. This discrepancy is rooted in the cosmological constant issue [1,28].

### 5.2. The Link Between Dark Energy and Vacuum Fluctuations

A proposition by Beck and Mackey [4] hints that equating the observed dark energy, driving the universe's accelerating expansion, to the cosmic vacuum energy density from theory implies a cut-off frequency for gravitationally active zero-point-energy vacuum fluctuations at around $\nu_c \cong 1.7$ THz. This suggests that only vacuum fluctuations with virtual photons below this frequency would be compatible with the relatively small observational data on dark energy.

### 5.3. Vacuum Fluctuations and Water Nanoclusters

Interestingly, this $\nu_c \cong 1.7$ THz frequency coincides approximately with the cutoff vibrational frequencies computed for magic-number n = 21,20 water nanoclusters (Figures 1 and 2a). As these clusters grow, their frequencies reduce, as depicted in Figures 2b,c and 3. In contrast, other cosmic molecules, including hydrogen, water monomers, carbon buckyballs, and organic molecules, do not exhibit pure vibrational cut-off frequencies in this THz range.

### 5.4 Rydberg Matter as a Quintessence Scalar Field and the Dynamical Casimir Effect

Viewing cosmic water nanoclusters as low-density Rydberg matter, they can potentially form a quintessence scalar field Q, characterized by an energy density, $\rho = ½\dot{Q}^2 + V(Q)$ [29]. This field's properties might be influenced by the absorption of vacuum fluctuations' virtual photons with frequencies surpassing $\nu_c \cong 1.7$ THz, as detailed through the microscopic dynamical Casimir effect [5]. This absorption mechanism transforms virtual photons into real ones, as depicted in Figure 8, ensuring that only the lower frequency photons exhibit gravitational activity. The absorbed photons can decay, releasing THz radiation as displayed in Figure 4.

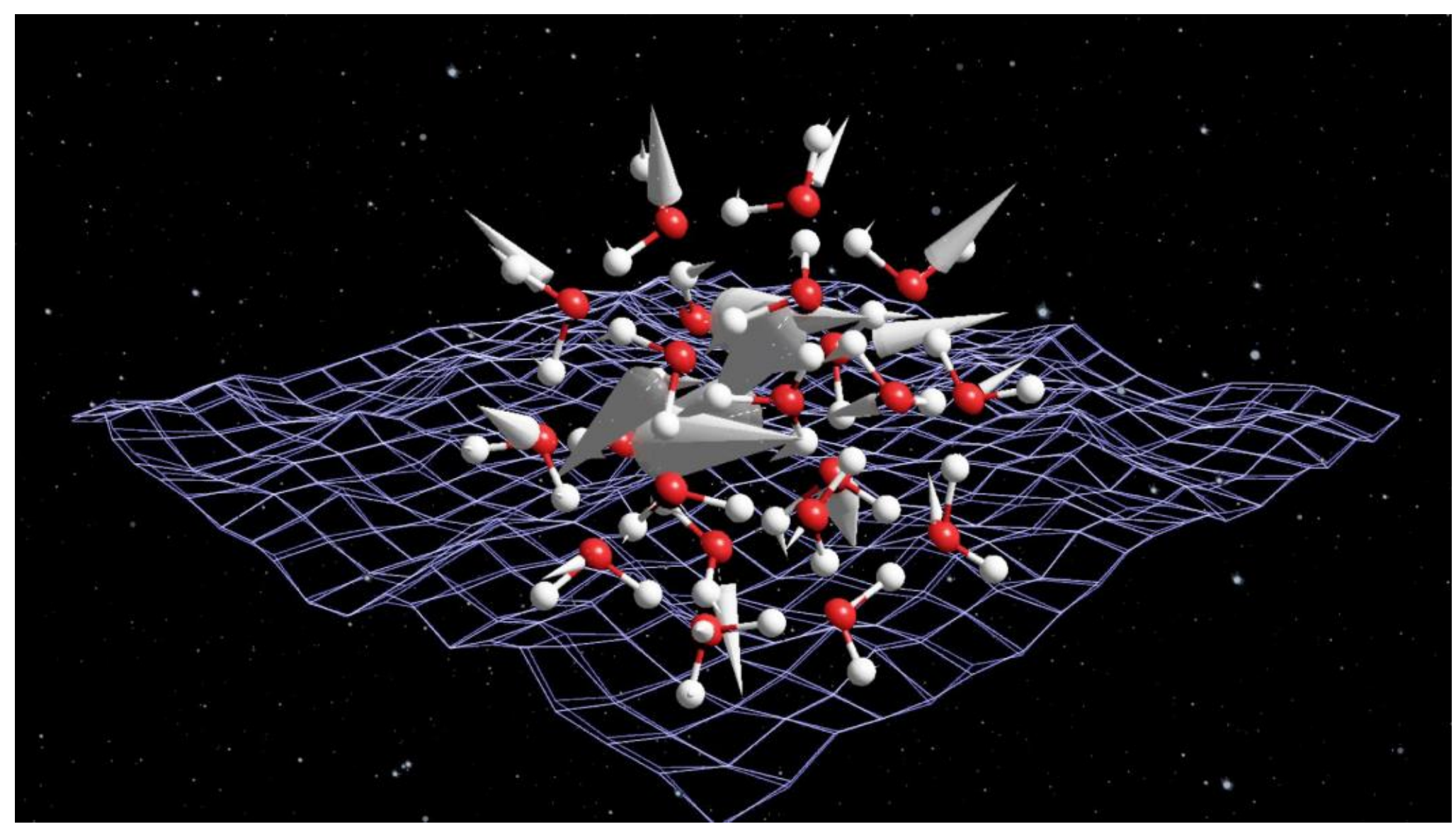

**Figure 8.** Dynamical Casimir absorption of vacuum fluctuation virtual photons by a cosmic water nanocluster, $\nu_c \cong 1.7$ THz vibrational mode.

**5.5 Cancellation of the Vacuum Energy Catastrophe**

To express the above scenario more formally, we traditionally regard the vacuum electromagnetic field (ignoring other fields) as a set of harmonic oscillators with normal-mode frequencies denoted by $\nu_{\boldsymbol{k}}$. By adding the zero-point energies of each of these oscillator modes, we arrive at the following energy density,

$$\rho_{vac} = \frac{E}{V} = \frac{1}{V} \sum_{\boldsymbol{k}} \frac{1}{2} h\nu_{\boldsymbol{k}} = \frac{4\pi h}{c^3} \int_0^\infty \nu^3 \, d\nu, \tag{1}$$

where $\boldsymbol{k}$ represents the normal modes of the electromagnetic field which align with the boundary conditions set by quantization volume, V. As V tends towards infinity, it corresponds to the right-hand side. To circumvent the diverging integral in Equation (1), one can introduce a cap with the cut-off frequency $\nu_c$ determined by the Planck scale, as suggested [1]. Nonetheless, this introduces a vast vacuum energy, which still overshadows the cosmologically observed value by many orders of magnitude. Alternatively, if we deduct from Equation (1) the energy density,

$$\rho_c = \frac{4\pi h}{c^3} \int_{\nu_c}^\infty \nu^3 \, d\nu \tag{2}$$

associated with the capture by cosmic water nanoclusters of vacuum-energy virtual photons above the nanocluster cutoff vibrational frequency, $\nu_c$, via the microscopic dynamical Casimir effect [5], the divergent integral in Equation (1) is predominantly negated. This results in the finite value presented in Equation (3), which can be associated with the dark energy density,

$$\frac{4\pi h}{c^3}\int_0^{\nu_c}\nu^3\,d\nu = \frac{\pi h\nu_c^4}{c^3} = \rho_{dark}. \tag{3}$$

The vibrational kinetic energy of small nanoclusters, represented as $½\dot{Q}^2$, is notably lower than their potential energy V(Q). This potential energy gets amplified to higher-THz-frequency "surface" vibrational modes (seen in Figures 1-3) when these nanoclusters capture vacuum photons, as illustrated in Figure 8. From this, we can infer that the quintessence scalar field pressure, given by $P = ½\dot{Q}^2 - V(Q)$ [29], becomes more negative as $\nu_c$ increases, and thus with the increase of $\rho_{dark}$. Planck satellite observations highlight that dark energy currently accounts for 68.3% of the universe's total known energy. This corresponds to a dark energy density, $\rho_{dark}$ = 3.64 GeV/m$^3$. From this, the cutoff frequency, $\nu_c$ is determined to be 1.66 THz, which aligns with the frequencies observed in the smallest pentagonal dodecahedral water clusters (Figures 1d and 2a). It's also worth noting that as water-cluster size grows (as shown in Figures 2b,c and 3), $\nu_c$ *decreases*. Consequently, if larger water clusters were to be increasingly ejected with time from cosmic dust, it suggests that the density of dark energy would *decrease* over time as per Equation (3). This could imply a *slowing expansion* of the universe over cosmic time.

## 6. The Cosmic Microwave Background

### 6.1 The CMB Spectrum

The prevailing belief in standard inflationary cosmology [30] posits that the universe's cosmic microwave background (CMB) spectrum emerged approximately 380,000 years post-Big Bang. Given the THz vibrational attributes of water nanoclusters addressed earlier, is there another potential contributor to the CMB that aligns with its spectrum? Historically, some theories have proposed that CMB might be due to thermal effects caused by "dust." This dust was thought to manifest as either hollow, spherical shells with a high dielectric constant [31] or as conductive needle-like grains [32]. It was suggested that radiation emanating from galaxy-sized entities at a redshift, $z \cong 10$, could be thermalized by a sparse concentration of such high-dielectric-constant dust. The large electric dipole moments of water nanoclusters (Figure 7a), in addition to their intense THz radiation emission [15], coincide with these criteria. Such water nanoclusters either form spherical "shells" of water-cluster O-H bonds (Figures 1&2) or needle-like aggregates (Figure 3). The amorphous layers of water-ice enveloping cosmic dust, from which these nanoclusters are derived, can be perceived as disordered nanoclusters possessing a high dielectric constant, and hence might also contribute. Astrophysical data revealed that the formation of early galaxies and reionization likely took place at redshifts, $z = 8.6$ and $z = 9.6$ [33,34]. This corresponds to approximately 600 million and 500 million years post-Big Bang, respectively. Additionally, more recent discoveries from the Hubble Space Telescope located a galaxy at $z \cong 11$, equivalent to approximately 400 million years after the Big Bang [35]. Even larger-redshift galaxies are being claimed by the Webb telescope [36]. When considering a redshift of $z \cong 10$, the distinct THz vibrational spectra of the water clusters are redshifted into the frequency domain of the observed CMB spectrum, more specifically near the peak of the leftmost CMB component shown in Figure 9. This suggests the THz emission from water nanoclusters originating around z ≅ 10, where the temperature is approximately 30K (thus allowing the existence of such clusters), might compete with the Big Bang CMB contribution originating from the earlier "recombination" phase at z ≅ 1100, where temperatures were around 4000K.

### 6.2 The CIRB Spectrum

The Cosmic Infrared Background (CIRB) is the light radiated from stars within innumerable dim galaxies. When one accounts for emissions from our solar system and the Milky Way, the CIRB spectrum shown in Figure 9 persists. The near-infrared region with wavelengths of approximately 2-3 microns, corresponds to starlight that has red-shifted in the infrared region. However, some starlight is also assumed to be absorbed by cosmic dust, which then emits it again in the far-infrared region, approximately 100 μm. However, it is proposed in this paper that cosmic dust, enveloped by amorphous water-ice, is composed of and ejects into space water nanoclusters, as depicted in Figures 1d and 2a. The water clusters present at z = 10 not only emit THz radiation that is redshifted to the leftmost CMB peak in Figure 9, but also, since they should exist in our Milky Way galaxy, contribute to the foreground radiation between approximately one and six THz in Figure 9. Indeed, the peak intensity shown around 1.5-1.7 THz may be associated with the dodecahedral water-nanocluster vibrational cutoff frequencies indicated in Figures 1d and 2a. This is the same maximum emission frequency of water clusters produced from water vapor in a laboratory vacuum, as shown in Figure 4 [15].

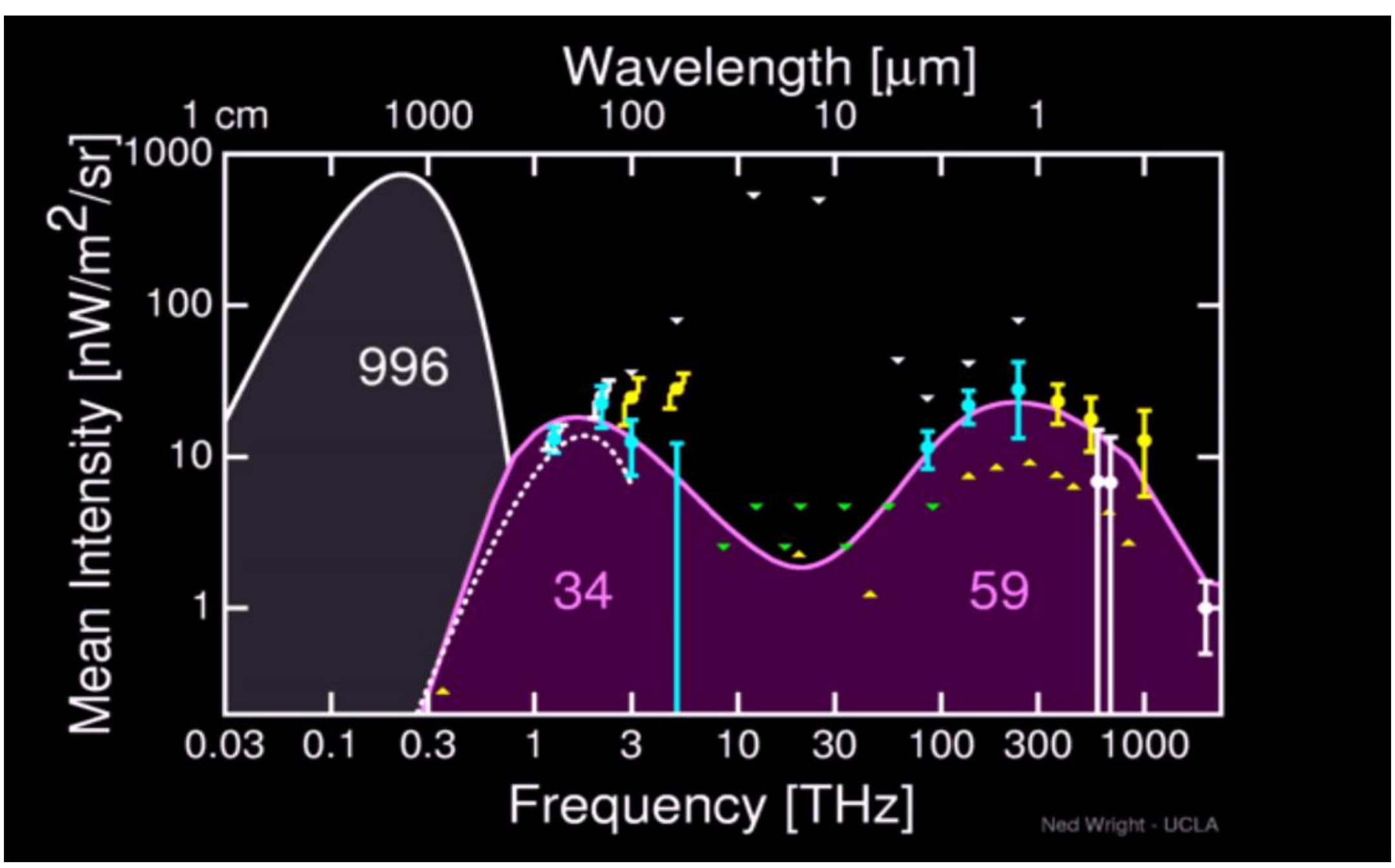

**Figure 9.** The Cosmic Infrared Background (CIRB) compiled by Ned Wright of UCLA. The data points correspond to a wide variety of observations.

The laboratory measurements displayed in Figure 4 demonstrate that when water nanoclusters emit THz radiation and are subjected to different ejection pressures in a vacuum chamber, the emission peaks change. Specifically, as the pressure decreased, both the frequency and intensity of the emission decreased. This corresponds to an increase in the size of the water clusters, as illustrated in Figures 2 and 3. The strongest emission, at 1.7 THz, is linked to water clusters of "magic numbers" n = 21 and 20 (Figure 4). The peaks that reduce in frequency and intensity down to roughly 0.5 THz can be attributed to larger clusters, as shown in Figures 2b,c and 3a. Notably, the smaller the water nanocluster, the higher the frequency of its emitted radiation, peaking near 1.7 THz. Drawing a comparison with cosmic scenarios, if we consider the radiation from water nanoclusters ejected from water-ice-covered cosmic dust at around $z \cong 10$, the redshifted emission from smaller versus larger clusters qualitatively resembles the CMB power spectrum. Further insights from the laboratory data [15] (Figure. 4), when connected with the ejection of water nanoclusters from cosmic dust, hint at a "pressure wave" originating around $z \cong 10$, which is about 500 million years post the Big Bang. This bears similarity to the CMB acoustic wave from the "recombination" period at $z \cong 1100$, approximately 380 thousand years after the Big Bang. In simpler terms, if we consider the observed CMB power spectrum to be even partly due to the redshifted THz radiation from cosmic water nanoclusters, it suggests a possible "turning point" for a *cyclic* expansion of our universe (see Section 9), in contrast to the inflationary Big Bang scenario originating from a singular quantum point. An interesting fact to ponder is that $z \cong 10$, or about 500 million years after the alleged Big Bang, aligns with the speculated period of the earliest (Population III) star formation. Most of these theorized stars would have short lifespans and quickly become supernovae which explode to cosmic dust that ejects water nanoclusters, which then act as coolant for rapid star creation (see Section 7), offering an explanation of the growing Webb telescope evidence for early galaxy formation [36].

#### 6.3 CMB Polarization: E- and B-Modes

Recent Planck satellite observations ascribed to foreground cosmic dust polarization in our galaxy have revealed that the power of the E-mode polarization is approximately twice that of the B-mode [37], although they are expected to be equal. Understanding this fact may be key to experiments looking for *primordial* B-modes leftover by inflationary gravitational waves, such as the infamous Antarctic BICEP2 experiments of 2014, which were dominated by foreground dust polarization after initial claims of primordial B-modes as possible support for cosmic inflation [38]. The E/B-mode polarization ratio can be explained not by the dust alone, but by the polarization of the water nanoclusters that comprise and are ejected from the amorphous water-ice enveloping the dust particles. This is due to the power of the three water-cluster $H_g$ "squashing" vibrational modes compared to the power of the two $H_u$ "twisting" modes represented in Figure 10, where the power is proportional to the squares of their amplitudes. They vibrate and twist over a 1-6 THz frequency range, possibly contributing redshifted radiation to the CMB and CMBIR.

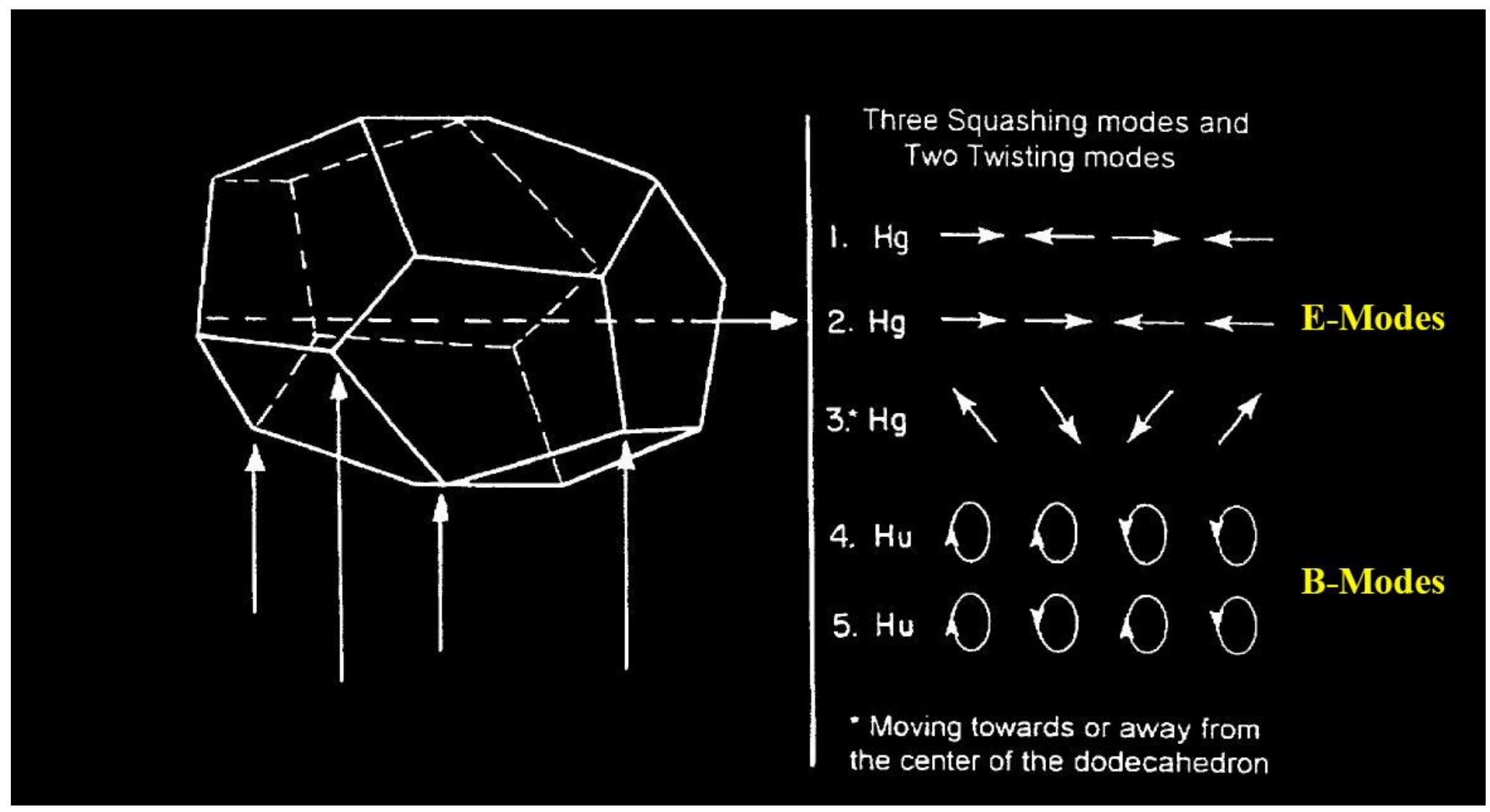

**Figure 10.** E- and B- vibrational modes of a pentagonal dodecahedral water nanocluster.

## 7. Cosmic Water Nanoclusters and Early Star Formation

Water vapor has been identified as a cooling agent in the process of star formation [8]. A remarkable finding revealed a vast water reservoir within a high-red-shift ($z \cong 4$) quasar, which had a water vapor mass surpassing 140 trillion times the combined water of the Earth's oceans and was 100,000 times more massive than the sun [10]. This, coupled with the suggestion [11] that water vapor might have been prevalent roughly a billion years post-Big Bang, leads to the hypothesis that there might have been a considerable presence of water nanoclusters approximately 500 million years after the Big Bang or at redshift $z \cong 10$. The rapid rate of star formation from dense gas clouds observed at $z \cong 6.4$, which is approximately 850 million years after the Big Bang [39], might be better understood considering the potential role of these water nanoclusters in aiding the cooling effects of the cloud's water vapor. Research on the infrared absorption of water nanoclusters, both in a controlled environment and within Earth's atmosphere, has shown that these clusters possess remarkable heat storage capabilities [12,40]. This property is attributed primarily to the *librational modes* of these nanoclusters, especially those around the frequency of 32 THz (1060 cm-1), as depicted in Figures 1 and 2. Consequently, the cooling process in star-forming gas clouds could potentially be driven by the emission of photons related to the mainly-oxygen vibrations on the cluster surface, falling within the 1–6 THz spectrum, as illustrated in Figures 1 and 2. Intriguingly, these photons might influence the CMB. In other words, if we accept the idea that THz radiation from spaceborne water nanoclusters around $z \cong 10$ plays a role in the CMB spectrum, then the CMB might also offer clues about the early stages of star formation.

## 8. Hubble Tension

The current model possibly explains the *Hubble tension,* where the Hubble constant value derived from observing supernovae suggests a faster expanding universe compared to values from CMB readings immediately after the Big Bang. It posits that cosmic dust enveloped by amorphous water-ice, which ejects water nanoclusters linked to dark energy and the acceleration of the universe, stems from stellar evolution. Given that the birth of the initial stars occurred after the reionization phase that succeeded the hydrogen formation recombination phase starting approximately 380,000 years post-Big Bang, substantial cosmic dust and, subsequently, cosmic water nanoclusters would not form until significantly later. This theory is bolstered by a recent sighting of the earliest cosmic dust 200 million years after the emergence of the first stars [41]. Over the universe's 13.8-billion-year expansion, cosmic dust volumes rose with more stars and galaxies emerging, culminating in today's sufficient dust and ejected water-nanocluster Rydberg matter to justify the recent value of the Hubble constant from supernovae observations.

## 9. Cosmic Water Nanoclusters and a Cyclic Universe

Inflationary cosmology suggests that in an expanding universe, dark matter's density diminishes more rapidly than dark energy's. This implies that dark energy will eventually take precedence, leading to a future where all cosmic matter is progressively dispersed by this expansion – a scenario often termed the "big rip." On the other hand, if we consider thar cosmic water nanoclusters, which emerge from the water-ice-encrusted cosmic dust as Rydberg matter, are analogous to a time-varying quintessence scalar field, as discussed in Section 5.4, our model introduces a potential cyclical relationship between dark matter and dark energy in the cosmos:

(1) As the universe expands and stars produce heavier elements, more amorphous water-ice-coated cosmic dust will arise from an increased number of supernovae. This dust increasingly releases larger water clusters owing to the expanding space and lower pressure, as shown in Figure 4. Figures 1-3 indicate that larger water clusters will have a decreasing vibrational cutoff frequency, $\nu_c$.

(2) Over time, as larger water clusters become more common, their vibrational cutoff frequency will tend towards $\nu_c \cong 0.5$ THz (see Figures 3 and 4) and likely even lower. This leads to a much-reduced dark energy density, $\rho_{dark}$ (see Equation 3), nearing zero, which implies a slowing expansion of the universe.

(3) Eventually, the gravitational pull from the remaining cosmic matter, resulting from fewer supernovae, dominates. The universe will cease its expansion, start contracting, and revert to a possible state such as 500 million years post Big Bang, where it is roughly 10% of its current size.

(4) When this happens, the pressure from the anticipated water-nanocluster vapor around $z \cong 10$ will rise. As per Figure 4, smaller water nanoclusters, particularly with magic numbers n = 21,20 and higher $\nu_c$ (seen in Figures. 1d and 2a), will become more prevalent. These clusters will serve as cooling agents for star formation (Section 7), leading once again to supernovae and the production of cosmic dust that produces water nanoclusters.

(5) This scenario provides a potential "turning point" for *non-inflationary re-expansion*, though further collapse and water vapor decomposition are possible to offer hydrogen for star development. These cycles, influenced by supernovae patterns, span billions of years, although the exact durations are uncertain. This *cyclic universe* model aligns qualitatively with others [42,43] and challenges the multiverse concept, which is a derivative of inflation theory.

## 10. Discussion

While this study ventures beyond prevalent elementary-particle, inflationary-cosmology and multiverse landscapes, it is not the sole divergent view. Other unified dark-matter-dark-energy scenarios [6,7] and cyclic-universe theories [42,43] have been proposed. This paper presents a consolidated approach that bridges known astrophysics with computational quantum astrochemistry. The notion that water nanoclusters dispersed in space constitute invisible Rydberg baryonic dark matter does not eliminate the possible existence of nonbaryonic dark-matter particles, such as WIMPS and AXIONS, although their observational evidence remains elusive [44]. As for dark energy, this study suggests that cosmic water nanoclusters can be described as a dynamical *quintessence* scalar field permeating space and absorbing the excess vacuum energy density predicted by quantum electrodynamics, as described in Equations (1-3). This view is compatible with the proposed dark matter "web" pervading the universe [45]. Additionally, the finding of a neutral hydrogen bridge between the Andromeda (M31) and Triangulum (M33) galaxies [46] suggests that cosmic water nanoclusters might be dispersed similarly, acting as Rydberg dark matter. The remarkable alignment of “magic-number” water-nanocluster THz vibrational frequencies with vacuum-energy THz frequencies, that lead to a dark-energy density consistent with cosmological data, might be coincidental. Nevertheless, considering a shared origin for both dark matter and dark energy outside conventional particle physics—a field yet to definitively pinpoint the sources of these phenomena—is enticing. Coincidentally, the magic-number nanocluster masses are the same order of magnitude as the lowest values estimated for WIMPS. No other known baryonic substances, including hydrogen, organic molecules, and fullerene buckyballs [47], share all these attributes to qualify as dark matter. Celestial entities such as planets, moons, asteroids, and gaseous nebulae harboring water-vapor and water-ice must also be considered potential water-nanocluster sources. For example, water clusters have recently been detected in the hydrothermal plume of Enceladus – a moon of Saturn [48].